\documentstyle[aps,multicol,psfig]{revtex}
\begin{document}

\title{Expansion and Contraction of Avalanches
in 2D Abelian Sandpile}

\author{D.V.Ktitarev$^{1,}$\footnote{Alexander von Humboldt
Research Fellow}$^,$\footnote{Permanent address: 
Laboratory of Computing Techniques,
JINR, Dubna,
141980 Russia} 
and V.B.Priezzhev$^2$}

\address{$^1$ Theoretical Physics, FB 10,
Gerhard-Mercator-University, 47048 Duisburg, Germany \\
 $^2$Laboratory of Theoretical Physics,
Joint Institute for Nuclear Research, Dubna,
141980 Russia }
\date{\today}

\maketitle

\begin{abstract}

  We present a detailed analysis of large scale simulations
  of avalanches in the 2D Abelian  sandpile model. We compare
  statistical properties of two different decompositions
  of avalanches into clusters of topplings and waves of
  topplings. Auxiliary   critical exponents are introduced
  and the existence of the exponent governing the contraction
  of avalanches claimed in our previous work
  [Priezzhev {\it et al}, PRL {\bf 76},2093 (1996)] is confirmed.
  We also give more elaborated argumentation for the exact values of
  the exponents characterizing the statistics of waves.
\vspace{.5cm}
\vspace{1.cm}
\end{abstract}
\begin{multicols}{2}
\section{Introduction}

The sandpile model introduced by Bak, Tang and Wiesenfeld \cite{BTW}
serves not only as a lapidary formulation of basic principles of
self-organized criticality (SOC) but also seems to be an appropriate
candidate for exact determination of all important critical exponents.
Indeed, the first steps following Dhar's discovery \cite{D} of the
Abelian structure of the sandpile model were encouraging.
They include determination of the total number of allowed configurations
in the recurrent state \cite{D}, evaluation of the height probabilities
\cite{H,P} and height-height correlation
functions \cite{H,Iv},
interpretation of the inverse Laplacian operator $\Delta^{-1}$ as an
expected number of topplings at a given site due to a particle added
to another one \cite{D}. Nevertheless, all analytical results obtained
up to now catch either static properties of the recurrent state or
diffusion-like dynamics of individual particles. The avalanche dynamics
as such, responsible for SOC, slips off the analytical description even
in the simplified limiting case  of large avalanches. The main obstruction
is that existing renormalization group methods \cite{Piet} neglect essential
peculiarities of the toppling process, and the complicated spatio-temporal
structure of avalanches prevents exact evaluation of the critical exponents.

To advance the analysis of avalanche dynamics, various decompositions of
avalanches in the Abelian sandpile model (ASM) into more
elementary objects have been proposed. In particular, Grassberger and Manna
noticed \cite{GM} that each avalanche can be represented
as a set of embedded clusters of sites related to a given number of topplings.
To make use of this construction for determination of critical exponents it
is desirable to obtain a dynamical procedure which
naturally divides the avalanche into a collection of clusters. It means that
due to the Abelian property of toppling operators, one can try to
change the order of topplings
so that each avalanche would expand to the largest cluster and then
contract by smaller and smaller sets of toppling sites.
In our previous works \cite{IKP94,IKP,PKI}
we made such an attempt proposing
a decomposition of avalanches into waves of topplings.

The main feature of the wave structure of avalanches is a possibility
to set up a one-to-one correspondence between waves and
two-rooted spanning trees \cite{IKP94}. Using the spanning
tree representation for waves, one can apply
the methods of graph theory to calculate the critical
exponents of wave and avalanche statistics \cite{IKP,PKI}.

On the other hand, it has been found out \cite{DM,IKP94}
that the set of waves and the set of clusters
for a particular avalanche do not coincide. Namely, waves
have such an irrelevancy in their
superimposing that the next wave can overlap the previous one
and the package of waves does not form embedded sets of sites like
clusters.
The observations of Dhar and Manna \cite{DM} and our simulations
on small lattices raise hope that the overlappings of waves
are relatively rare events.  We have assumed \cite{PKI} that one can neglect
the difference between clusters and waves of topplings  and
consider each next wave embedded into the previous one as a typical
situation. Based on this assumption we suggested
a method of evaluation of the basic critical exponents of 2D ASM.
Our latest simulations, however, have shown that the next
wave typically overlaps the previous one. Moreover, the large-scale
simulations of Paczuski and Boettcher \cite{PB} state that
the average difference of sizes between two subsequent waves
is actually negative.

Nevertheless, we will show in this paper that it is possible
to modify our simplified scenario of the avalanche process and to
describe the phases of expansion and contraction in terms of the
wave decomposition. Besides, we will demonstrate
that the theoretical predictions are in complete agreement
with the numerical data obtained by Paczuski and Boettcher \cite{PB}.

The paper is organized as follows. In Section II we formulate
the ASM model, define avalanche clusters   and
waves of topplings and introduce the basic ideas of expansion and
contraction of avalanches. Section III is devoted to the description
of the local dynamics of waves and proof of the existence of the
contraction exponent. In Section IV we present analytical
derivation of the exponents of conditional distribution of waves
obtained by Paczuski and
Boettcher \cite{PB} from extensive numerical simulations.
In Section V, an elucidative point of view
on the renormalization group approach \cite{Piet} to the sandpile model is
suggested.

\section{basic concepts}

We remind the definitions of the model, waves of topplings,
avalanche clusters and explain our basic ideas.

In 2D ASM one
starts from the empty square lattice (occupation numbers $z_i=0$ for
all sites) and drops sand, particle by particle, at random sites:
$z_i \rightarrow z_i+1$. If any $z_i > 4$, the site $i$ is unstable and
topples: $z_j \rightarrow z_j - \Delta _{ij}$ where $\Delta$ is the
Laplacian matrix. The toppling at $i$ may cause instability at its nearest
neighbors.
The subsequent topplings continue until there are no more
unstable sites. Then, one adds again a particle at random site
initiating a new chain of topplings and so on. The process of toppling
during each perturbation is called an avalanche and the set of toppled
sites form a compact cluster of all toppled sites.

To obtain the wave decomposition of an avalanche  \cite{IKP},
one has to topple all sites that become unstable after adding a
particle at $i$ keeping this site
out of the second toppling. The set $W_1$ of toppled sites is the
first wave of topplings.  All sites except maybe the site $i$
become stable after the first wave. If the resulting height $z_i > 4$,
one topples the site $i$ the second time and continues the relaxation process
not permitting this site to topple the third time. The new set $W_2$ of
relaxed sites is the second wave. The process continues until the site $i$
becomes stable and the avalanche stops.

Grassberger and Manna \cite{GM} defined clusters $C_n, n=1,...,M$
of sites toppled not less than $n$ times during the given avalanche,
$M$ is the number of topplings at the initially perturbed site.
The sets $C_n, n \le M$ are all compact and each $C_n$ contains the clusters
$C_{n+1},...,C_M$.

It is possible to evaluate the asymptotics of cluster size
distribution considering the set of generated clusters without
reference to a particular avalanche they belong to.
According to \cite{D},
the expected number of topplings at site $j$ due to adding a particle at
site $i$ is given by the lattice Green function $G_{ij}=[\Delta^{-1}]_{ij}$.
The number of topplings $G_{ij}$  coincides with the expected number
of clusters $C_n$ containing the site $j$ in an avalanche started at
$i$. Therefore, the probability that the linear extent $r$ of a cluster
exceeds the distance $|i-j|$ between $i$ and $j$ is
\begin{equation}
Prob(r>|i-j|) \sim  G_{ij},
\label{1}
\end{equation}
Using the known asymptotics of the Green function for large distances
$G(r)\sim \ln (r)$ and compactness of clusters
(cluster area $s_c\sim r^2$), we get
\begin{equation}
P(s_c)= P(r)\frac{dr}{ds_c} \sim \frac{1}{s_c}.
\label{2}
\end{equation}

It was established in \cite{IKP94}, that every wave is a compact set
without holes and each site in a wave topples exactly once in that wave.
Thus, the expected number of topplings $G_{ij}$ given by the lattice
Green function can be expressed alternatively by the probability that a
wave taken from an arbitrary avalanche initiated at site $i$ covers site
$j$. Writing Eq.( \ref{1}) for waves, we get again the size distribution
similar to Eq.(\ref{2})
\begin{equation}
P(s_w) \sim \frac{1}{s_w},
\label{3}
\end{equation}
where $s_w$ is the area of a wave.

To find critical exponents characterizing the size
distribution of  avalanches,
one also needs a general picture of the avalanche process as a whole.
In the case of clusters, the picture is quite clear. The set of
clusters is ordered and each next cluster is embedded into the
previous one. However, the clusters of topplings, being convenient
for a computer decomposition of avalanches, are hardly reproducible
by dynamical rules as each cluster grows monotonically during the
whole avalanche process.

On the contrary, the wave construction admits a simple dynamical
interpretation but loses the property of ordering
which is inherent in the case of clusters. In spite of the
irregularity of waves, we are still able to use a partial
ordering of waves assuming that a typical
avalanche consists of two phases:
fast expansion and slow
contraction. The first phase contains relatively few waves with a large
negative difference between subsequent waves
$\Delta(s_k)=s_{k}-s_{k+1}$.
The second phase forms the main body of an avalanche with a positive average
difference $\Delta(s_k) > 0$. In \cite{PKI} the fast phase was
associated with
the single first wave which reaches at once a maximal size the given
avalanche spreads. The positive difference
for the rest of the waves was assumed
to be dependent only on the size of a preceding wave and satisfied the
scaling law
\begin{equation}
\langle \Delta (s) \rangle \sim s^{\alpha}
\label{4}
\end{equation}
for large $s$.

\begin{figure}
\centerline{\psfig{figure=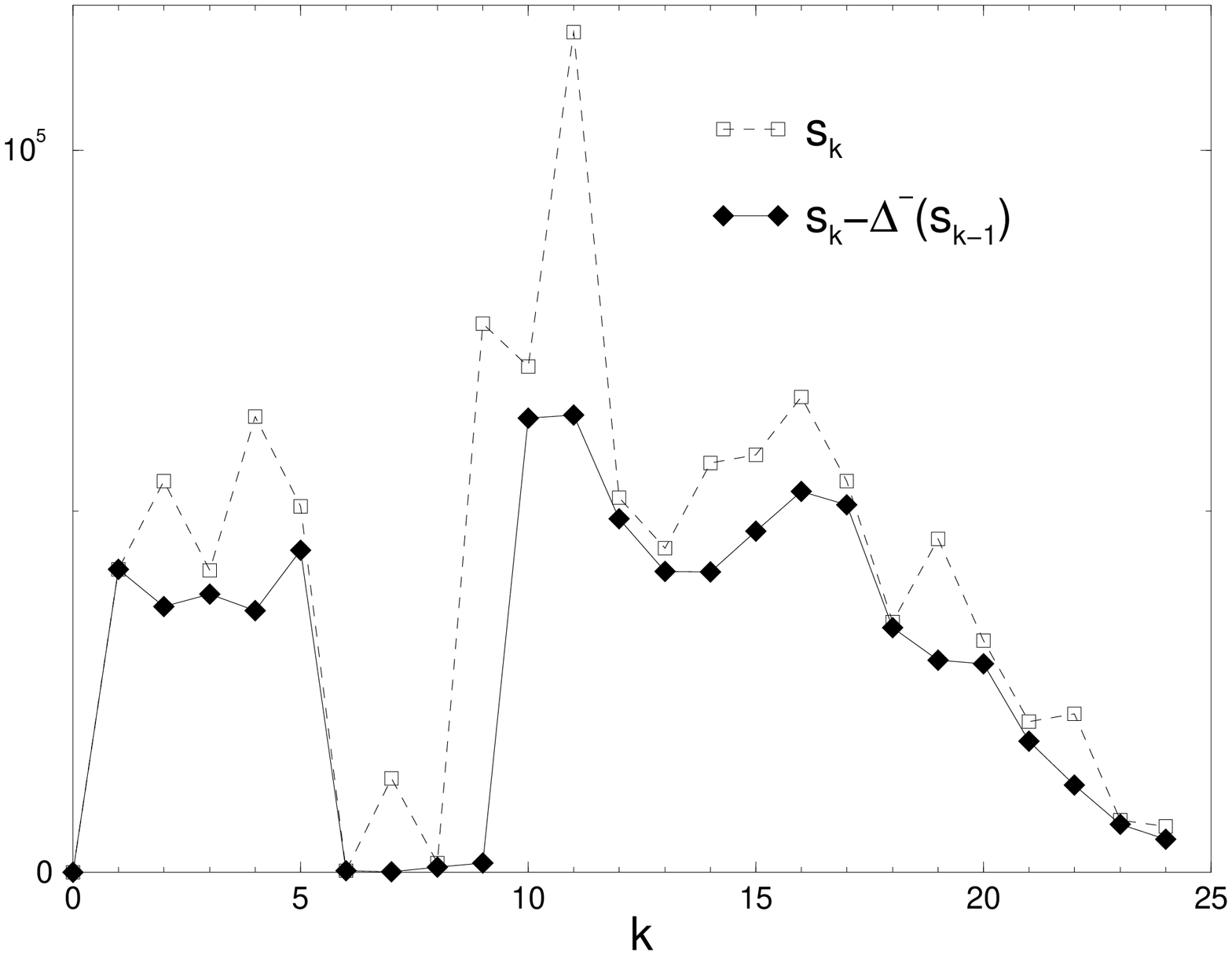,width=7cm,angle=-90}}
\label{fig1}
{\small
FIG. 1. Sizes $s_k$ of waves in a typical avalanche on the lattice
of size $L^2$, $L=500$ (empty squares); the same quantities subtracted
by the size of overlapping: $s_k-\Delta^-(s_{k-1})$ (filled diamonds).
}
\end{figure}

If the law Eq.(\ref{4}) is valid for clusters of
topplings as well, the density of clusters can be defined as the average
number of  clusters of size between $s_c$ and $s_c + ds_c$ in one avalanche
of the size $S > s_c$
\begin{equation}
\frac{dn}{ds_c}=\frac{1}{s_c^{\alpha}}.
\label{5}
\end{equation}
By assumption, the density depends on $s_c$ but not on $S$.
Then, the critical exponent $\tau$ in the distribution of a number of
sites covered by an avalanche $P(S) \sim S^{-\tau}$ can be related with
the exponent $\alpha$ in Eq.(\ref{4}). Indeed, the probability
distribution of cluster sizes $P(s_c)$ is proportional to the
probability of avalanches whose size $S$ exceeds $s_c$:
$P(S>s_c)\sim s_c^{-\tau+1}$ and to the density of clusters Eq.(\ref{5}):
\begin{equation}
P(s_c)\sim s_c^{-\tau+1}s_c^{-\alpha}.
\label{6}
\end{equation}
Comparing Eq.(\ref{6}) with Eq.(\ref{2}) we obtain
\begin{equation}
\tau + \alpha = 2,
\label{7}
\end{equation}
and the problem of finding the basic exponent $\tau$ is reduced to search for
the exponent $\alpha$ which is related more directly to details of the
avalanche process.

The 'contraction' exponent $\alpha$ is well defined
for avalanche clusters or for waves provided that one can neglect
the differences between these two kinds of objects.
In this connection, the following questions arise.
Is it possible to define the 'contraction' exponent for waves
taking into account overlappings and if so,
what is a correspondence between its numerical values for clusters and waves?
Can we establish the same relation Eq.(\ref{7}) for waves and use their
spanning tree structure to estimate the critical exponents of the model?

In the following sections we discuss these questions using large scale
simulations of clusters
of topplings, a more elaborated analysis of waves and new numerical data
for subsequent waves obtained by Paczuski and Boettcher \cite{PB}.

\begin{figure}
\centerline{\psfig{figure=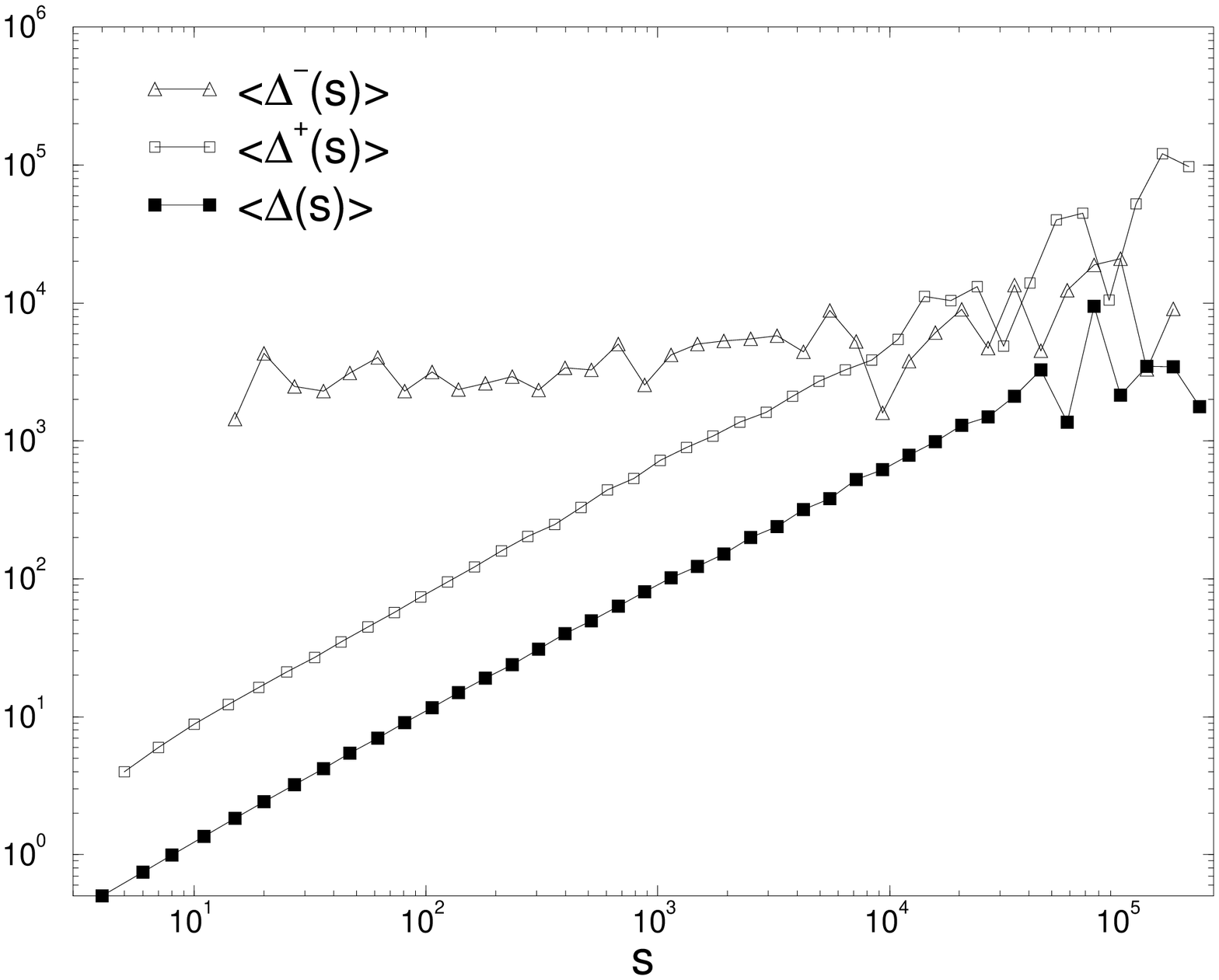,width=7cm,angle=-90}}
\label{fig2}
{\small
FIG. 2.  The average values of $\Delta^-(s), \Delta^+(s)$
for waves and $\Delta(s)$ for clusters as functions of their size $s$
(see text for definitions),
obtained from the simulations data
of $10^6$ avalnches on lattice of size $L=500$. The graph
$\Delta(s)$ is shifted vertically.
}
\end{figure}

\section{contraction exponent for waves}

First, we present a picture of a typical avalanche of 2D
ASM on a square lattice of size $L^2$ for $L=500$. On Fig.1
we plot the size $s_k$ of the wave as a function of its number $k$
in the avalanche. We can see that many of the next waves
have the size greater than the size of the previous one. Moreover,
even those waves, whose size is actually less than the
size of its predecessor, are not most frequently  embedded
into the set of sites formed by the previous wave. For
the particular avalanche presented on Fig.1 the event of
overlapping the previous wave by the next one occurs for
all waves except the sixth and the last one.

Anyway, one can note that a typical avalanche contains several sharp
peaks corresponding to fast expansion of the avalanche size and
in-between intervals of relatively slow, although irregular,
contraction. One may expect that the average difference between
subsequent waves in the slow phase follows a scaling law similar to
Eq.(\ref{4}). To verify this, one has to extract from the averaging
of $\Delta(s_k)$ those waves which are related
to the expansion phase.
We can avoid this cumbersome and ambiguous procedure introducing
new variables characterizing the 'local' contraction and expansion.

Consider two typical subsequent waves of topplings
$W_k$ and $W_{k+1}$ with the sizes $s_k$ and $s_{k+1}$, the $(k+1)$st wave
overlaps the $k$th wave. Let $W$ be their intersection
having the size $s$.

Define the variables $\Delta^+(s_k)=s_k-s$ and
$\Delta^-(s_k)=s_{k+1}-s$,
the first quantity is 'local contraction', the second one
refers to 'local expansion'. We calculated  the averages
$\langle \Delta(s_k) \rangle$ for clusters,
$\langle \Delta^+(s_k) \rangle$ and $\langle \Delta^-(s_k) \rangle$
for waves of topplings
using data of $10^6$ avalanches for the system size $L=500$ (Fig.2).
The simulations show a power-law behavior
$\langle \Delta(s_k) \rangle \sim s^\alpha$
for clusters and $ \langle \Delta^+(s_k)\rangle \sim s^{\alpha^+}$ for waves;
the exponents $\alpha$ and $\alpha^+$ have close values.
The value of the exponent $\alpha^-$,
for the relation
$\langle \Delta^-(s_k) \rangle \sim s^{\alpha^-}$
is much smaller than $\alpha^+$.

Concerning the estimation of the exponents
$\alpha$, $\alpha^+$ and $\alpha^-$, we have
to point out that the numerical determination of these
values is a rather difficult problem because of a slow convergence
of data obtained for large lattices to their limiting values.
So, our numerical results
$\alpha \approx \alpha^+\approx 0.88$ and
$\alpha^- \approx 0.29$
for $L=500$ are still far from the expected limit.
The problem of estimation of these exponents is somehow similar to
the numerical determination of the exponent
$\tau$ (for discussion, see, for example, \cite{LU}).
The extrapolation  $L \rightarrow \infty$ gives us some wide
interval of possible values of $\alpha$ and $\alpha^+$
which includes 3/4, the theoretical prediction \cite{PKI}
for the exact value.

Being equivalent to $\tau$ from a computational point of view,
the exponent $\alpha^+$ is more convenient for theoretical evaluations.
The spanning tree representation of waves \cite{IKP94} makes it possible
to interpret $\Delta^{+}(s)$ as a sum of branches attached to the boundary
of a wave \cite{PKI}, and then to use exact results obtained for
the one-component Potts model \cite{SD}.

The relative magnitude of $\Delta^-(s_k)$ and $\Delta^+(s_k)$
is such that for large $s$
the contraction of avalanche dominates its expansion.
We show on Fig.1 by filled diamonds the sizes of waves $s_k$ subtracted
by the size $\Delta^-(s_{k-1})$. It is clear that neglecting the
quantities $\Delta^-(s_k)$ we do not change the qualitative
dynamical picture of the avalanche and the contraction of waves
can be described in terms of $\Delta^+(s_k)$.
Based on these data we can also see that the average
$\langle\Delta(s_w) \rangle$ for waves which is equal to the remainder
$\langle \Delta^+(s_k)\rangle-\langle\Delta^-(s_k)\rangle$
is actually negative for small waves and positive for large ones,
as it was found in \cite{PB}.

Finally, we establish the relationship
of the type of Eq.(7) for the exponents $\tau$ and $\alpha^+$.
Following argumentation for clusters (section II),
we estimate the known asymptotics of size distribution $P(s_w)$ of waves
Eq.(\ref{3})
which is proportional to the
probability of avalanches whose size $S$ exceeds $s_w$:
$P(S>s_w)\sim s_w^{-\tau+1}$ and to the density of waves
\begin{equation}
P(s_w)\sim s_w^{-\tau+1}\frac{dn}{ds_w}.
\label{3.2}
\end{equation}
Let $N(s,s-t)$ be the number of waves in a particular avalanche with
sizes between $s-t$ and $s$ provided that the size of the given
avalanche is greater than $s$.
The asymptotic behavior of $dn/ds_w$ can be evaluated as
\begin{equation}
\frac{dn}{ds_w} \sim \frac{\langle N(s_w,s_w-t)\rangle}{t}
\label{3.3}
\end{equation}
for large $s_w \gg t$.
Take in Eq.(\ref{3.3}) $c\langle\Delta^+(s_w)\rangle$
instead of $t$ where $c=O(1)$ is
a constant. Then, from Eq.(\ref{3.2}) we obtain
\begin{equation}
s_w^{-1} \sim s_w^{-\tau-\alpha+1}\langle 
N(s_w,s_w-c\langle\Delta^+(s_w)\rangle)\rangle.
\label{3.4}
\end{equation}

\begin{figure}
\centerline{\psfig{figure=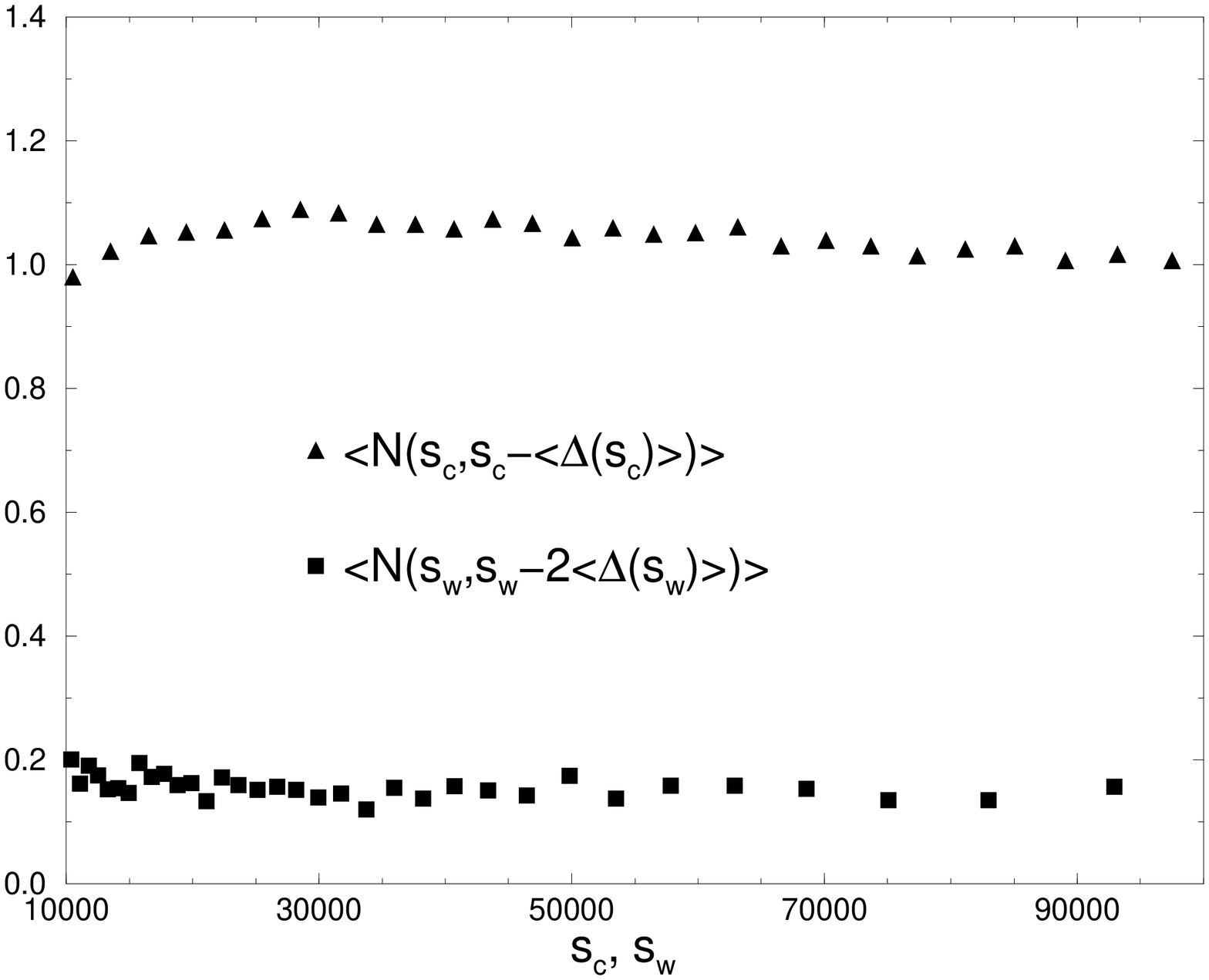,width=7cm,angle=-90}}
\label{fig3}
{\small
FIG. 3. The expected number $\left<N(s,s-t)\right>$ of clusters
(waves) of size from the interval $[s-t,s]$ provided that the size of
the corresponding avalanche is greater than $s$.
Due to fluctuations the data for waves are uncertain for some
large values of $s_w$, so the averages are not proportional. 
}
\end{figure}

On Fig. 3 we present the quantity
$$
\langle N(s_w,s_w-2\langle\Delta^+(s_w)\rangle)\rangle
$$ as a
function of $s_w$ calculated from our numerical data where we put
$c=2$.
It is apparently evident that asymptotically
it scales as $O(1)$. Thus, we obtain from Eq.(\ref{3.4})
\begin{equation}
\tau + \alpha^+ = 2.
\label{3.5}
\end{equation}
For comparison on Fig.3 we present the
function
$$
\langle N(s_c,s_c-\langle\Delta(s_c)\rangle)\rangle
$$
for clusters which
also scales as $O(1)$ confirming Eq.(7).

\section{analysis of conditional wave distribution}

Recently, Paczuski and Boettcher \cite{PB} have undertaken careful
numerical simulations to find the size distribution of subsequent waves for a
given size of the preceding wave $P(s_{k+1}|s_k)$. The data was
represented by a scaling form
\begin{equation}
P(s_{k+1}|s_k) \sim s_{k+1}^{-\beta}F(\frac{s_{k+1}}{s_k}),
\label{2.1}
\end{equation}
where $F(x) \rightarrow const$ when $x \rightarrow 0$ and
$F(x \gg 1) \sim x^{-r}$.
The function $P(s_{k+1}|s_k)$ being considered as a normalized distribution
of $s_{k+1}$ should be multiplied by the factor $s_k^{\beta-1}$
to provide the normalization condition
\begin{equation}
\int\limits^{s_{co}}P_N(s_{k+1}|s_k)ds_{k+1} = \int\limits^{s_{co}/s_k}
x^{-\beta}F(x)dx,
\label{2.2}
\end{equation}
where $s_{co}\sim L^2$ is the cutoff in the wave sizes from the finite
system size \cite{PB}. The normalized function $P_N(s_{k+1}|s_k)$ has the
asymptotics
\begin{equation}
P_N(s_{k+1}|s_k) \sim s_{k+1}^{-\beta}s_{k}^{\beta-1},
\label{2.3}
\end{equation}
when $s_{k+1} \ll s_k$, and
\begin{equation}
\label{2.4}
P_N(s_{k+1}|s_k) \sim s_{k+1}^{-\beta-r}s_{k}^{\beta+r-1},
\end{equation}
when $s_{k+1} \gg s_k$.

To be consistent with the analysis in \cite{PKI}, both exponents
$\beta$ and $r$ should be explained from the same point of view based
on the spanning tree representation of waves. Before doing that, we
will discuss an attempt to verify the existence of the scaling law
Eq.(\ref{4}) for the contraction phase of an avalanche by calculating
the average difference between subsequent waves
$\langle \Delta(s_k)\rangle = \langle s_{k}-s_{k+1} \rangle$ \cite{PB}.
By the assumption \cite{PKI},
the average $\langle \Delta(s_k) \rangle \sim s_k^{\alpha} > 0$
for the main part of
an avalanche corresponding to the process of slow contraction of wave
fronts. A serious problem, however, is how to select the waves relating
to the slow phase. In a real avalanche, at least waves
corresponding to the largest contribution to the expansion should be
removed from the averaging as it was explained in Section II.
Without the selection of waves responsible for
the contraction of avalanches   the result of
averaging of $\Delta s$ obtained in \cite{PB} is easily predictable.

Following \cite{PB}, fix a value $s$ and take all waves with $s_k=s$
together with the subsequent waves of size $s_{k+1}$ from all avalanches whose
size $S \geq s$. Consider separately the cases $s_{k+1} < s_k$ and
$s_{k+1} > s_k$. In the first case, the average difference
$\langle \Delta s^{(-)} \rangle = \langle s_k - s_{k+1} \rangle$
is obviously $\langle \Delta s^{(-)} \rangle < s $.

In the opposite case, the waves with $s_{k+1} > s_k$ have a power law
asymptotics $P(s_{k+1}) \sim s_{k+1}^{-\theta}$
where  $1 < \theta < 2$ for all $s_k$.
Indeed, the size distribution of waves with an arbitrary origin
is $P(s) \sim s^{-1}$ according to the two-component
spanning tree representation \cite{IKP}. The distribution of waves
of size $s$ with the origin in a
fixed unique site is $P'(s) \sim s^{-2}$.
In the considered case, the subsequent wave starts at a site in the
localized area inside the previous wave. This implies $1 < \theta < 2$.
Therefore, the averaged positive difference
$\langle \Delta s^{(+)}\rangle = \langle s_{k} - s_{k+1}\rangle
 \sim L^{2(2-\theta)} - s$ and diverges
with the lattice size $L$. Thus, $\langle \Delta s^{(+)}\rangle >
\langle \Delta s^{(-)}\rangle $
until $s$ becomes large and finite-size effects become essential.
We see that the negative
values of $\langle \Delta s(s_k) \rangle $ obtained in \cite{PB}
are not actually surprising and indicate only that the simple average
$\langle \Delta s \rangle $ does not exhibit a power-law dependence
on $s_k$ and cannot be related directly to the density of waves.

Nevertheless, the distribution Eq.(\ref{2.1}) itself brings important
information on avalanches. The exponents characterizing its asymptotics
are related to basic exponents of the sandpile model.

Consider first
the exponent $\beta$. According to Eq.(\ref{2.1}), this exponent determines
the behavior of smaller waves following just after waves of larger sizes:
$s_{k+1} \ll s_k$.
The wave $W_k$ can be represented by a tree $T_k$
covering the area $s_k$ and having the root $i$ at the point where the
wave $W_k$ was initiated \cite{IKP}. The tree $T_{k+1}$ representing the wave
$W_{k+1}$ has the root at the same point $i$.

\begin{figure}
\centerline{\psfig{figure=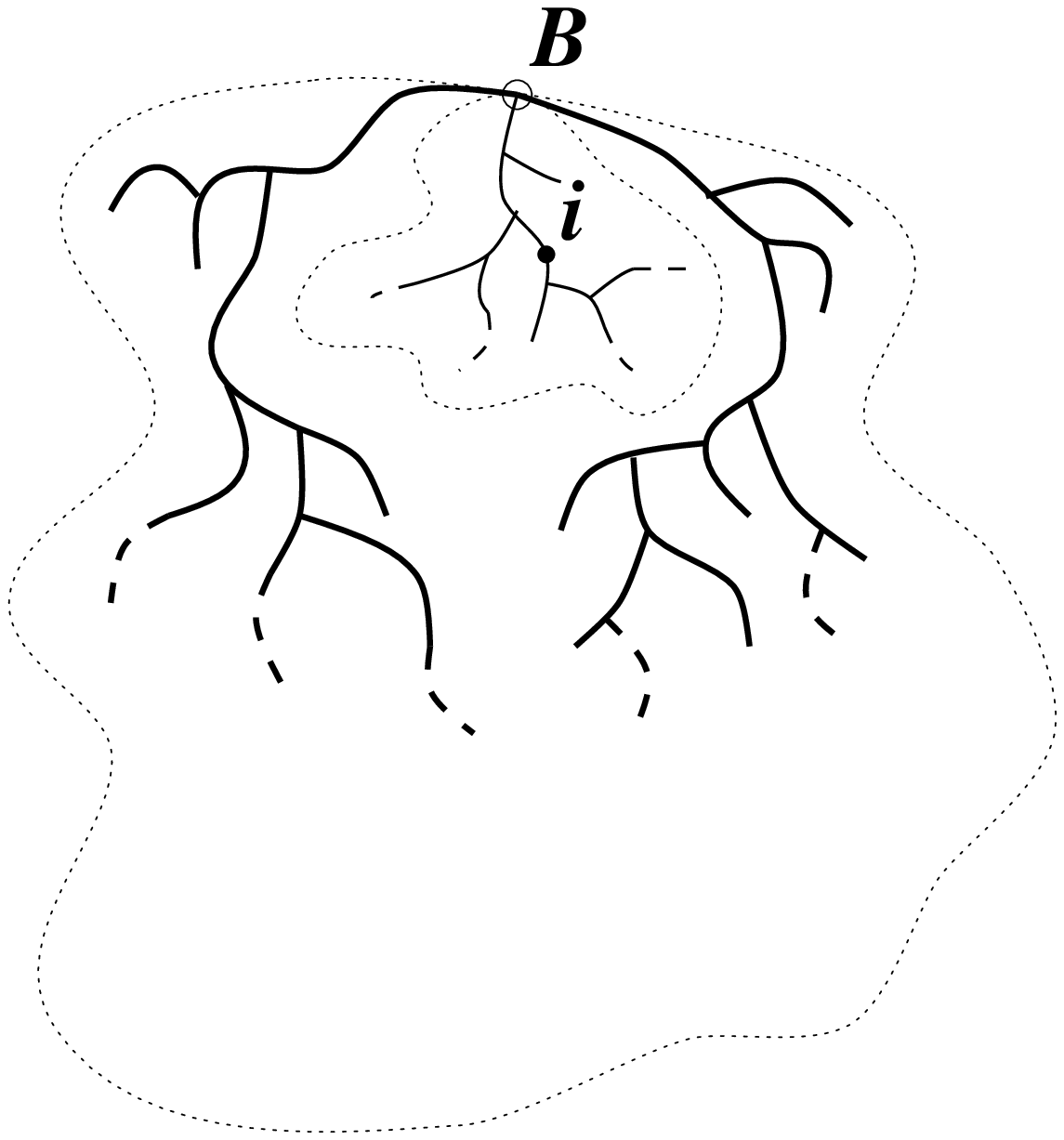,width=7cm,angle=-90}}
\label{fig4}
{\small
FIG. 4. The structure of spanning trees representing two subsequent
waves with sizes $s_{k+1} < s_k$.
The origin of the avalanche
is denoted by $i$. $B$ is the point of intersection of the
boundaries of these waves.
}
\end{figure}

To provide  the sharp
reduction of the next wave, $T_k$ and $T_{k+1}$ must have a special structure.
Since each site in a wave topples exactly once, the state of the system
inside a wave does not change after the wave is completed. This means that any
branch of $T_{k+1}$ attached to the root $i$ repeats a branch of $T_k$
until it ends inside $s_k$ or touches the boundary of $s_k$.
It follows from
$s_{k+1} \ll s_k$
that at least one point $B$ exists where the boundaries
of $W_k$ and $W_{k+1}$ intersect (Fig. 4). In the vicinity of $B$, both
trees  $T_k$ and $T_{k+1}$ can be attached by a bond $b$ to their complemented
subtrees $T_k^{'}$ and $T_{k+1}^{'}$ \cite{IKP} defined by the condition that a
unification $T$ and $T^{'}$ gives a complete spanning tree of the whole
lattice. Inversely, deletion of the bond $b$ from the corresponding
trees produces subtrees $T_k$ and $T_{k+1}$ which obey the statistics of
disconnected branches
of lattice spanning trees \cite{MDM} or, equivalently, the asymptotics of
last waves \cite{IKP94}
\begin{equation}
P_{last}(s) \sim \frac{1}{s^{11/8}}.
\label{last}
\end{equation}
It has been
demonstrated in \cite {PB} that the function $F(x)$ in Eq.(\ref{2.1})
is constant in a finite interval $0<x<c<1$. Thus, we can consider the
distribution
\begin{equation}
P_c(s_{k+1}) = \int \limits_{s_k>s_{k+1}/c} P_N(s_{k+1}|s_k)ds_k
\label{2.5}
\end{equation}
with the function $ P_N(s_{k+1}|s_k)$ taken in the form Eq.(\ref{2.3}).
This gives
\begin{equation}
P_c(s_{k+1}) =s_{k+1}^{-\beta} \int \limits_{s_k>s_{k+1}/c}
s_k^{\beta-1}ds_k \sim \frac{L^{2\beta}}{s_{k+1}^{\beta}}.
\label{2.6}
\end{equation}
On the other hand, $ P_c(s_{k+1})$ apart from the normalization factor
is given  by a joined probability distribution of disconnected branches
 $T_k$ and $T_{k+1}$. Despite the fact that subtrees   $T_k$ and $T_{k+1}$
are strongly connected ($T_{k+1}$ is a part of $T_k$), the distributions
of their sizes can roughly be considered  as independent. Then, we obtain
\end{multicols}
\vspace{.2cm}
\begin{equation}
P_{last}(s_{k+1})P_{last}(s_k>s_{k+1}/c) =s_{k+1}^{-11/8}
\int \limits_{s_k>s_{k+1}/c} s_k^{-11/8}ds_k \sim \frac{1}{s_{k+1}^{7/4}}
\label{2.7}.
\end{equation}
\vspace{.2cm}
\begin{multicols}{2}
To get $ P_c(s_{k+1}) $ from Eq.(\ref{2.7}), we have to multiply the
last expression by $s_{k+1}$, the number of possible positions of the
root $i$ inside the disconnected branch $T_{k+1}$. Finally,
$P_c(s_{k+1}) \sim 1/s_{k+1}^{3/4} $ and comparing with
Eq.(\ref{2.6}) we get $\beta=3/4$ which explains the numerical
result \cite{PB}.

To relate the exponent $r$ with
the exponent $\tau$ in the distribution of a number of distinct sites
covered by an avalanche, we consider waves of two types. A wave
will be referred to as the growing one or the G-wave if $s_{i+1} \geq s_i$
and the reducing one or the R-wave if $s_{i+1} < s_i$. Every avalanche
corresponds to a unique sequence of G-waves and R-waves e.g. GRGGR....

The number of distinct sites $s_d$ in an avalanche is proportional to
the size of the maximal wave
$W_{max}$, so we can expect that
\begin{equation}
P(s_{max}) \sim \frac{1}{s_{max}^{\tau}}
\label{2.8}
\end{equation}
with the same critical exponent $\tau$.

The expected number of  waves in an avalanche  diverges
logarithmically with the lattice size \cite{H}. However, if the
idea about the fast expansion phase is correct, the expected
number of waves in the interval between the  maximal wave
and the latest wave
$W_{k_0}$ with $s_{k_0} \sim O(1)$ before the maximal wave
should be finite when $L \rightarrow \infty$.

Starting with this assumption, consider a finite
sequence of $n$ G-waves and R-waves between the waves
$W_{k_0+1}$ and $W_{k_0+m}=W_{max}$ (for simplicity, we denote
their sizes by $s_1,...,s_n$).
The first and the last waves in the sequence are
clearly of type G. It follows from the numerical data of \cite{PB}
that the asymptotics Eqs.(\ref{2.3}) and  (\ref{2.4}) of the
distribution function $P_N(s_{k+1}|s_k)$ are factorized. Extrapolating
the distribution  Eq.(\ref{2.4}) to the case of vanishing previous
waves
$s_{k_0} \rightarrow 1$,
we can obtain the distribution of first
waves in the sequence:
\begin{equation}
P(s_{1}) \sim \frac{1}{s_{1}^{\beta+r}}.
\label{2.9}
\end{equation}

For an avalanche GG... beginning from two G-waves, the distribution
of the second wave is given by
\begin{equation}
P(s_{2})=
\int \limits^{s_2}ds_1 P(s_1)P_N(s_2|s_1).
\label{2.10}
\end{equation}
Using Eqs.(\ref{2.4}) and (\ref{2.9}) we have for large $s_2$
\begin{equation}
P(s_{2}) \sim \frac{ln s_2}{s_{2}^{\beta+r}}.
\label{2.11}
\end{equation}
Similarly, the leading asymptotics for the n-th wave in a sequence
of n G-waves is
\begin{equation}
P(s_{n}) \sim \frac{(ln s_n)^{n-1}}{s_{n}^{\beta+r}}.
\label{2.12}
\end{equation}
The presence of R-waves reduces the logarithmic divergence of the
numerator. For instance, using Eq.(\ref{2.3}) and Eq.(\ref{2.4}) we
get in the case GRG... the numerator $ln s_3$ instead of $(ln s_3)^2$
in the case GGG... .
Generally, if  $k$ $(k \leq n-2)$ last G-waves in the sequence follow
R-wave, the convolution
\end{multicols}
\begin{equation}
P(s_{n})=
\int...\int ds_1...ds_{n-1} P(s_1)P_N(s_2|s_1)...P_N(s_n|s_{n-1})
\label{2.13}
\end{equation}
\begin{multicols}{2}
has the asymptotics
\begin{equation}
P(s_{n}) \sim \frac{(ln s_n)^{k}}{s_{n}^{\beta+r}}.
\label{2.14}
\end{equation}
Thus, for any finite sequence of G-waves and R-waves between the
relatively small $k_0$-th wave
and the maximal wave, we have the leading exponent $\beta+r$ which governs the
distribution of the maximal waves $P_{max} \sim s_{max}^{-\tau}
\sim s_{max}^{-\beta-r}$.
The numerical values obtained in \cite{PB} are $\beta = 3/4, r = 1/2$.
This gives $\tau = 5/4$ obtained in \cite{PKI} from scaling arguments.
\section{Discussion}

In conclusion, the analysis of the decomposition of avalanches into waves
of topplings shows that a difference between two subsequent
waves can be described by appropriate variables which follow a power
law dependence on the wave size $s$. The exponent $\alpha^{+}$
corresponding to the contraction of waves can be related to one of
the basic avalanche exponents $\tau$.

The relation between the asymptotics Eq.(\ref{2.4}) of the distribution
of subsequent waves $P_N(s_{k+1}|s_k)$ in the case  $s_{k+1} \gg s_k$
and the exponent $\tau$  in the distribution of distinct sites
involved into an avalanche implies an alternative way of determining
$\tau$. Instead of derivation $\tau$ from the analysis of slow
contraction process, we can use the statistics of large waves $W_{k+1}$
overlapping their predecessors $W_k$ to link $\tau$ with the exponents
$\beta$ and $r$. This approach sheds new light on the renormalization
group (RG) procedure proposed by Pietronero {\it et al.} \cite{Piet}
for the sandpile model. In the RG method, one deals with sites of three
classes: stable, critical and unstable. Extending the characterization
of the stationary properties at a generic scale, one describes the
propagation of instability through the lattice taking into account
only one-shot relaxation events at each scale. Thus, proliferation
effects due to multiple relaxations are not considered in this scheme.
In this respect, the process described by RG is not a true avalanche,
rather it is a wave propagating from a given point or from a cluster
of a given size. Correspondingly, the critical exponent determined
in this way is actually the sum of exponents $\beta + r$ in the
asymptotics of distribution of large waves. Its numerical value 1.248
obtained in \cite{Zhenya} is in excellent agreement with the value
$\beta + r = 5/4$ proposed in \cite{PB}. On the other hand, it was
shown in Section IV  that  $\beta + r = \tau$ which explains the
validity of the RG approach despite the neglect of multiple
relaxations.

\vskip 1cm
\centerline{\small \bf ACKNOWLEDGEMENTS}

This work was partially supported by the Russian Foundation  for
Basic Research through Grant No. 97-01-01030.
V.B.P. appreciates  hospitality of the Computational Physics Group
at Duisburg University.


\end{multicols}
\end{document}